\documentclass{article}
\usepackage{spconf,amsmath,graphicx}

\title{DETECTING FALSE ALARMS AND MISSES IN AUDIO CAPTIONS}
\name{Rehana Mahfuz, Yinyi Guo, Arvind Krishna Sridhar, Erik Visser}
\address{Qualcomm Technologies Inc.}
\begin{document}
%
\maketitle
\begin{abstract}
Metrics to evaluate audio captions simply provide a score without much explanation regarding what may be wrong in case the score is low. Manual human intervention is needed to find any shortcomings of the caption. In this work, we introduce a metric which automatically identifies the shortcomings of an audio caption by detecting the misses and false alarms in a candidate caption with respect to a reference caption, and reports the recall, precision and F-score. Such a metric is very useful in profiling the deficiencies of an audio captioning model, which is a milestone towards improving the quality of audio captions.
\end{abstract}
\section{Introduction}
\label{sec:introduction}
The possibility of automatically describing an auditory scene using text is an exciting advancement in humanity's effort to improve awareness without the need for expensive human attention. While video monitoring deprives subjects of the scene from much-valued privacy while also being energy-hungry, audio monitoring is a reasonable choice for maintaining surveillance while also preserving privacy and being energy-efficient. Audio captioning, which is the task of describing audio using text, can enable a wide variety of solutions. In the industry, it can be used for machine condition monitoring, or for security systems. In personal lives, audio captioning can afford people peace of mind in the form of smart monitoring when they leave their dependent loved ones or pets at home.

In the advancement of audio captioning, one bottleneck has been the lack of a transparent method to evaluate the quality of the audio captions. Current methods to evaluate audio captions simply provide a score, without much of an explanation regarding what may be wrong with the caption. While hallucination detection and mitigation in text generation applications such as summarization have been considered, no such effort has been made for audio-to-text-generation. In this work, we introduce a method to identify mistakes in the caption. Specifically, our method automatically detects the false positives and false negatives in the candidate caption, with respect to the reference caption. To our knowledge, this is the first framework which automatically detects shortcomings of the caption, which is a first step towards developing strategies to address problems in the audio captioning model.

\section{Related Work}
\label{sec:related_work}
The framework of evaluating audio captions involves the availability of a reference caption, which is generally human-generated, based on which the quality of a candidate caption is determined.
Current methods to evaluate the quality of audio captions can be divided into three categories. BLEU \cite{bleu}, METEOR \cite{meteor} and ROUGE \cite{rouge} are borrowed from machine translation, and consider the overlap between words or matching between synonyms to establish similarity.
From image captioning, CIDER \cite{cider} considers the cosine similarities between Term Frequency-Inverse Document Frequencies (TF-IDFs) \cite{tfidf} of n-grams of the captions, while SPICE \cite{spice} determines the overlap between scene graphs created from the reference and candidate captions separately. 
Pre-trained language models are also being leveraged to judge semantic similarity between audio captions, such as in BERTScore \cite{bertscore}, Sentence-BERT \cite{sentence-bert}, FENSE \cite{fense} and SPICE+ \cite{spice+}. An effort has also been made to consider the time of occurrence of audio events to establish correspondence \cite{t2a-grounding-metric}.

\section{PROCEDURE}
\begin{figure}[h]
\includegraphics[width=8.5cm]{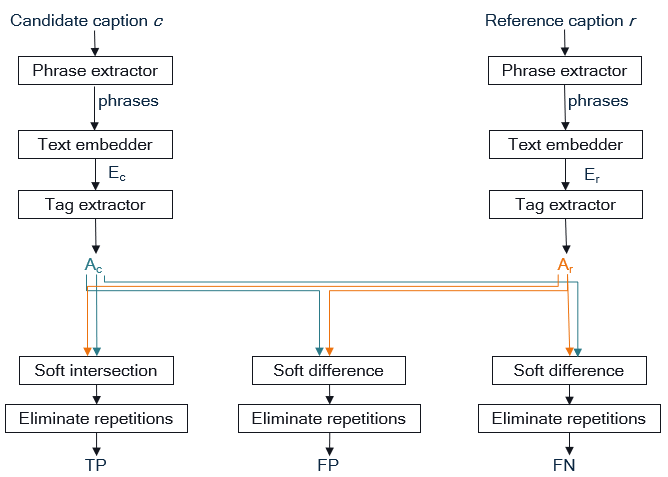}
\caption{Method of finding false alarm and misses in an audio caption.}
\label{fig:method}
\end{figure}

\label{sec:procedure}
We propose a method to obtain false positives and false negatives in a candidate caption $c$ from a reference caption $r$.
Let $A=\{a_1, a_2, ..., a_M\}$ be a nearly comprehensive set of $M$ audio classes, as contained in the dataset AudioSet \cite{audioset}, and let $E_A = \{txt\_emb(a_i), txt\_emb(a_2), ..., txt\_emb(a_M)\}$ be a set of text embeddings of these audio classes, obtained using some function $txt\_emb$.
From each caption $c$ and $r$, we first obtain phrases by matching their Parts-of-Speech (POS) tags with standard patterns of POS tags of phrases, using a function $phrases$. Then we get text embeddings of each phrase to obtain $E_c = \{txt\_emb(p) \forall p \in phrases(c)\}$ and $E_r = \{txt\_emb(p) \forall p \in phrases(r)\}$.
Next, we identify the collection of audio tags in the candidate caption by isolating their text embeddings in a set 
\begin{equation} \label{eq:ac}
A_c = \{t \in E_c: cos\_sim(t, u) > tag\_t \forall u \in E_a\} 
\end{equation}
, which is the set of text embeddings of all audio tags whose text embedding's cosine similarity with a candidate caption's phrase text embedding exceeds a threshold $tag\_t$, and $cos\_sim$ refers to the cosine similarity.
Similarly, we also calculate the set of text embeddings of audio tags in the reference caption as 
\begin{equation} \label{eq:ar}
A_r = \{t \in E_r: cos\_sim(t, u) > tag\_t \forall u \in E_a\}
\end{equation}
Since AudioSet's tag ontology has multiple entries with nuanced meanings for some categories such as \textit{Music} and \textit{Engine}, we eliminate redundancies in $A\_c$ and $A\_r$, where redundancy is defined by the existence of another element in the set with whom the cosine similarity exceeds a threshold $rep\_t$, as shown in Equation \ref{eq:eliminate_rep}.
Then, we use this information to identify the true positives by calculating the set of text embeddings of audio tags $TP$ as shown in Equation \ref{eq:tp}, which is the set of all members of $A_c$ whose cosine similarity with any member of $A_r$ exceeds a threshold $sim\_t$. This represents the audio captions captured by both the candidate caption and the reference caption.
Next, we identify the false positives by calculating the set of text embeddings of audio tags $FP$ as shown in Equation \ref{eq:fp}, which is the set of all members of $A_c$ whose cosine similarity with all members of $A_r$ is below the threshold $sim\_t$. This represents all the audio tags suggested by the candidate caption, but absent in the reference caption.
Similarly, we identify the false negatives by calculating the set of text embeddings of audio tags $FN$ as shown in Equation \ref{eq:fn}, which is the set of all members of $A_r$ whose cosine similarity with all members of $A_c$ is below the threshold $sim\_t$. This represents all the audio tags present in the reference caption, but not captured by the candidate caption.

\begin{equation} \label{eq:tp}
TP = \{t \in A_r: \exists u \in A_c: cos\_sim(t, u ) > sim\_t\}
\end{equation}
\begin{equation} \label{eq:fp}
FP = \{t \in A_c: \forall u \in A_r cos\_sim(t, u) < sim\_t\}
\end{equation}
\begin{equation} \label{eq:fn}
FN = \{t \in A_r: \forall u \in A_c cos\_sim(t, u) < sim\_t\}
\end{equation}

Equation \ref{eq:tp} can viewed as a soft version of the set intersection between $A_r$ and $A_c$, where the softness comes from the cosine similarity. Along those lines, Equation \ref{eq:fp} may be viewed as a soft version of the the set difference between $A_c$ and $A_r$. Similarly, Equaltion \ref{eq:fn} can be viewed as a soft version of the set difference between $A_r$ and $A_c$.

Before using the lengths of $TP$, $FP$ and $FN$ to calculate recall, precision and F-score, we again eliminate redundancies in each of these sets, as shown in Equation \ref{eq:eliminate_rep}. We propose this similarity-based F-score as a metric to evaluate the quality of audio captions, and abbreviate it as SBF, which stands for Similarity-Based F-score..

\begin{equation} \label{eq:eliminate_rep}
S=\{t \in S: \forall u \neq t \in S cos\_sim(t, u) < rep\_t \}    
\end{equation}

\section{EXPERIMENTS}
\label{sec:experiments}

\begin{table*}[]
    \centering
    \begin{tabular}{|l|l|l|l|l|l|l|l|l|l|l|}
    \hline
        tag\_t & 0.4&&&0.45&&&0.5&& \\ \hline
        AudioCaps & Precision & Recall & F-score  & Precision & Recall & F-score  & Precision & Recall & F-score \\ \hline
        Clotho & 0.425 & 0.228 & 0.249 &  0.378 & 0.297 & 0.284 & 0.341 & 0.303 & 0.282 \\ \hline
    \end{tabular}
   \caption{SBF scores on captions generated by a vanilla audio captioning model using the evaluation splits of AudioCaps and Clotho.}
    \label{table:basic_result}
\
\end{table*}

\begin{table*}[h]
    \centering
    \begin{tabular}{||p{0.15\linewidth}| p{0.35\linewidth}| p{0.3\linewidth}| p{0.15\linewidth}||}
    \hline
        & Caption & Phrases & Tags \\ \hline \hline
        Candidate caption & A bell is ringing while birds are chirping in the background & a bell is ringing; birds are chirping in the background & Bell; Bird \\ \hline
        Reference caption & A bell rings while people talk in a courtyard & a bell rings; people talk in a courtyard & Bell; Conversation \\ \hline
        True Positives & Bell & ~ & ~ \\ \hline
        False Positives & Bird & ~ & ~ \\ \hline
        False Negatives & Conversation & ~ & ~ \\ \hline \hline
        Candidate caption & The waves are crashing against the shore and splashing & the waves are crashing against the shore; splashing & Splash, splatter; Waves (surf); Water \\ \hline
        Reference caption & Ocean waves roll in and out from the shore & ocean waves roll in; out from the shore & Waves (surf); Ocean \\ \hline
        True Positives & Waves (surf); Ocean & ~ & ~ \\ \hline
        False Positives & - & ~ & ~ \\ \hline
        False Negatives & - & ~ & ~ \\ \hline \hline
        Candidate caption & Rain is pouring down the street with traffic sounds & rain is pouring down the street; traffic sounds & Rain; Traffic noise, roadway noise \\ \hline
        Reference caption & A river is flowing relatively swiftly and a waterfall flows & a river is flowing; a waterfall flows & Waterfall; Stream; Water; Raindrop \\ \hline
        True Positives & Raindrop & ~ & ~ \\ \hline
        False Positives & Traffic noise, roadway noise & ~ & ~ \\ \hline
        False Negatives & Stream & ~ & ~ \\ \hline
    \end{tabular}
    \caption{Examples of how false alarms and misses are detected.}
    \label{table:example}
\end{table*}

\begin{table*}[h]
    \centering
    \begin{tabular}{|l|l|l|l|l|l|l|l|l|l|}
    \hline
        ~ & ~ & AudioCaps-Eval & ~ & ~ & ~ & Clotho-Eval & ~ & ~ & ~ \\ \hline
        ~ & ~ &  HC &  HI & HM &  MM &  HC &  HI & HM &  MM \\ \hline
        all-MiniLM-L6-v2 & Sentence-BERT & 0.64 & 0.984 & 0.921 & 0.836 & 0.586 & 0.95 & 0.741 & 0.641 \\ \hline
        ~ & FENSE & 0.581 & 0.955 & 0.891 & 0.816 & 0.595 & 0.943 & 0.797 & 0.717 \\ \hline
        ~ & SBF & 0.409 & 0.935 & 0.921 & 0.664 & 0.529 & 0.898 & 0.638 & 0.574 \\ \hline
        paraphrase-TinyBERT-L6-v2 & Sentence-BERT & 0.64 & 0.988 & 0.925 & 0.73 & 0.6 & 0.955 & 0.759 & 0.673 \\ \hline
        ~ & FENSE & 0.645 & 0.98 & 0.916 & 0.85 & 0.605 & 0.947 & 0.802 & 0.731 \\ \hline
        ~ & SBF & 0.409 & 0.931 & 0.921 & 0.707 & 0.529 & 0.885 & 0.703 & 0.6 \\ \hline
    \end{tabular}
    \caption{Correlation of metrics with human judgments on the AudioCaps-Eval and Clotho-Eval datasets.}
    \label{table:fense_dataset_eval}
\end{table*}

\subsection{Qualitative evaluation}

We used our framework to evaluate captions generated by a vanilla audio captioning model. The model as in \cite{xinhao_paper}, which uses a CNN10 \cite{pann} encoder and a transformer decoder, was trained with the training split of the AudioCaps dataset \cite{audiocaps} for 30 epochs, and further fine-tuned with the training split of the Clotho dataset \cite{clotho} for 60 epochs. The batch size was 32 and learning rate was 0.001. The best checkpoint as judged by the SPIDER score on the validation split was used. We set $sim\_t$ and $rep\_t$ to 0.45. For all our experiments, to get text embeddings, we use the \textit{all-MiniLM-L6-v2} model of Sentence-BERT \cite{sentence-bert}, unless otherwise mentioned.
\subsection{Quantitative evaluation}

We leverage the availability of human judgments on pairs of audio captions indicating which one describes a given audio file better. The AudioCaps-Eval and Clotho-Eval datasets \cite{fense} provide such human judgments for 1,671 pairs from the AudioCaps dataset and on 1,750 caption pairs from the Clotho dataset. As performed in \cite{fense}, we measure the correlation of SBF's judgments with human judgments.
To obtain the text embeddings, we experiment with using \textit{paraphrase-TinyBERT-L6-v2}, which is 240 MB, and \textit{all-MiniLM-L6-v2} models, which is 80 MB.

\begin{figure}[]

\begin{minipage}[b]{0.48\linewidth}
\includegraphics[width=4.5cm]{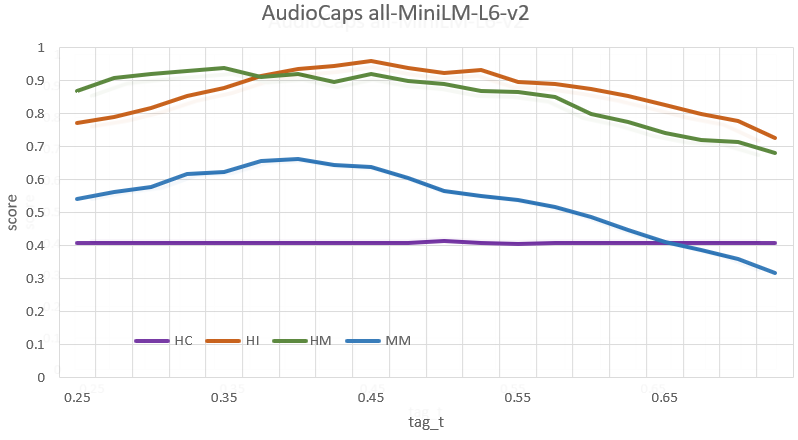}
\caption{AudioCaps-Eval, using \textit{all-MiniLM-L6-v2}}
\end{minipage}
\hfill
\begin{minipage}[b]{0.48\linewidth}
\includegraphics[width=4.5cm]{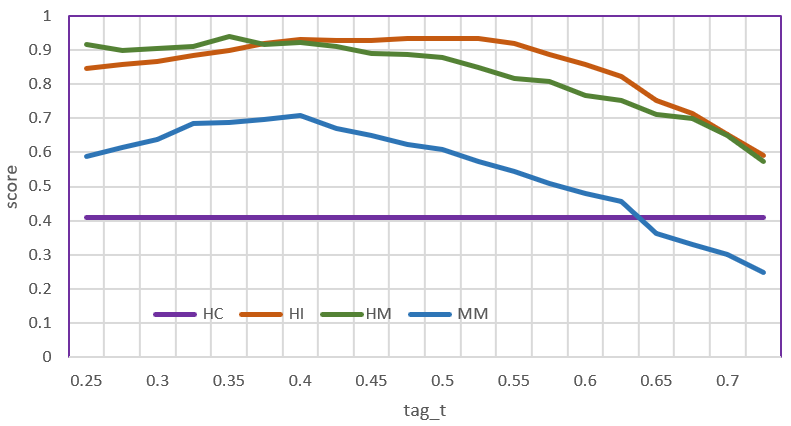}
\caption{AudioCaps-Eval, using \textit{paraphrase-TinyBERT-L6-v2}}
\end{minipage}

\begin{minipage}[b]{0.48\linewidth}
\includegraphics[width=4.5cm]{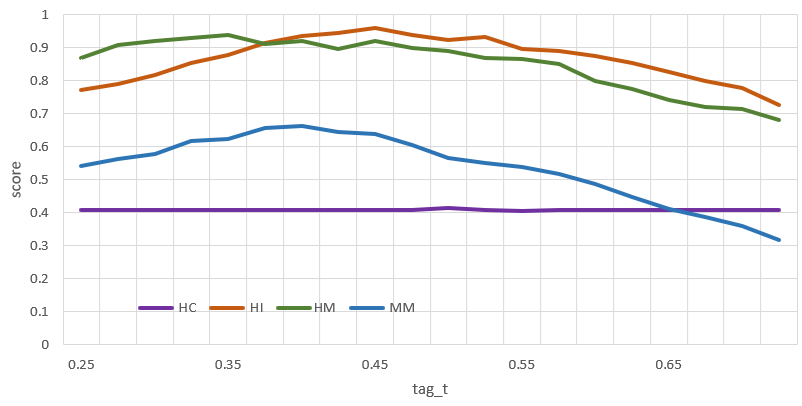}
\caption{Clotho-Eval, using \textit{all-MiniLM-L6-v2}}
\end{minipage}
\begin{minipage}[b]{0.48\linewidth}
\includegraphics[width=4.5cm]{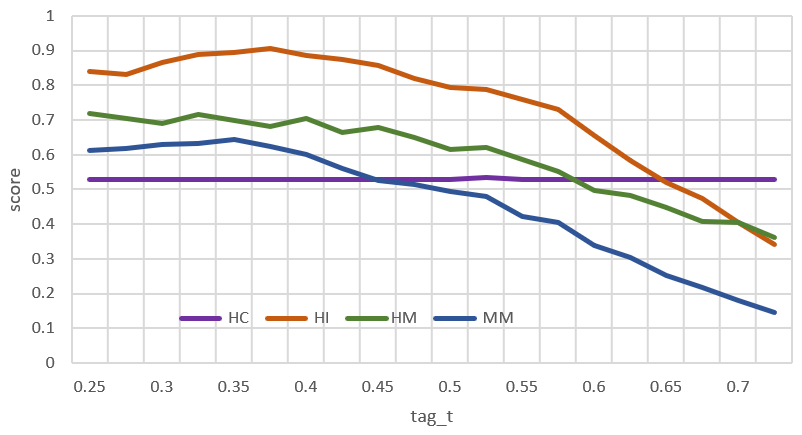}
\caption{Clotho-Eval, using \textit{paraphrase-TinyBERT-L6-v2}}
\end{minipage}

\caption{Correlation of SBF's judgments with human judgments, while varying $tag\_t$ with two different datasets and two different text embeddings.}
\label{fig:graphs}

\end{figure}

\section{RESULTS}
\label{sec:results}
\subsection{Qualitative results} \label{subsec:qualitative_results}

From the AudioCaps and Clotho evaluation splits, Table \ref{table:example} shows some examples of how false alarms and misses are detected.
In the first example, in the candidate caption, from phrases ``a bell is ringing" and ``birds are chirping in the background", tags \textit{Telephone bell ringing}, \textit{Doorbell}, \textit{Bell}, \textit{Church bell}, \textit{Bicycle bell}, \textit{Jingle bell}, \textit{Bird vocalization, bird call, bird song} and \textit{Bird} are detected. After eliminating repetitions, we are left with ``Bell" and ``Bird" as detections.
In the reference caption, from phrases ``a bell rings" and ``people talk in a courtyard", tags  \textit{Telephone bell ringing}, \textit{Bell}, \textit{Church bell}, \textit{Bicycle bell}, \textit{Doorbell}, \textit{Jingle bell}, \textit{Tubular bells} and \textit{Conversation} are detected. After eliminating repetitions, we are left with \textit{Bell} and \textit{Conversation}.
Detection of multiple tags related to \textit{Bell} shows the importance of our repetition eliminator.
By intersecting the sets of tags obtained from the candidate and reference captions, we get $Bell$ as a True Positive.
By subtracting the set of candidate tags from the set of reference tags, we get \textit{Bird} as a False Positive.
By subtracting the set of reference tags from the set of candidate tags, we get \textit{Conversation} as a False Negative.
The second example illustrates the case of a perfect scenario, where both the candidate ad reference captions mention the same or similar sounds.
In the third example, \textit{Stream} is detected as a False Negative even though \textit{Rain} and \textit{Stream} are both related to \textit{Water}, because the reference caption mentions a river, and the candidate caption doesn't. If we want to be more lenient and not count this as a False Negative, we would have to decrease $sim\_t$.

Table \ref{table:basic_result} shows the precision, recall and F-scores obtained using the evaluation splits of the AudioCaps and Clotho datasets. If we increase $tag\_t$, more audio tags are detected, which increases recall, but decreases precision. Increasing $sim\_t$ would make our evaluation framework more sensitive to variations in meaning, since it would raise the threshold required for tags to be considered similar. Similarly, increasing $rep\_t$ would also increase sensitivity to variations in meaning.
\subsection{Quantitative results} \label{subsec:quantitative_results}

Figure \ref{fig:graphs} shows how our our metric's judgment's correlation with human judgments varies as we vary \textit{tag\_t}. A value of 0.4 for \textit{tag\_t} seems reasonable to achieve good correlation with human judgment.
Hence this value of \textit{tag\_t} is used in Table \ref{table:fense_dataset_eval}, which shows the performance of different metrics with the AudioCaps-Eval and Clotho-Eval datasets.
Since SBF is primarily designed to detect mistakes, it does not perform well on HC, HM and MM, because in pairs belonging to these categories, both captions are correct. We can see that SBF performs better on HI, because one caption in the pair is indeed incorrect.
By comparing the use of two text embedding models, where one is three times the size of the other, we see a noticeable difference in the quality of their judgments in some cases.
\section{CONCLUSION}
\label{sec: conclusion}


We propose a novel method to detect mistakes in an audio caption in the form of false alarms and misses. Having these detections provides insights into the deficiencies of a model which generated the audio captions. Often, false alarms result from over-representation of certain sounds in the training data, such as \textit{Snoring}, \textit{Horse trotting} and \textit{Spray}. Sometimes these are also cross-triggers. Similarly, misses are actually cross-triggers in disguise (example: \textit{Spray} instead of \textit{Vehicle}). Understanding shortcomings of a system is a first step towards remediation measures, such as adjusting training data or adjusting training strategies, which will be explored in the future..

\section{DISCUSSION}
\label{sec: discussion}
While this work provides a way to detect false alarms and misses, and also to rule out cross-triggers caused by semantically similar sounds such as \textit{Water} and \textit{Rain}, it does not rule out cross-triggers caused by groups of sounds which are acoustically similar but semantically different, such as \textit{Vehicle} and \textit{Spray}, or \textit{Frying} and \textit{Rain}. This could be addressed by considering acoustic similarity between sounds based on their audio embeddings, instead of only semantic similarities based on their text embeddings.
Another direction enabled by our framework is to forgo reliance on the availability of audio captions, and instead use audio tags obtained from a audio tagging model. The challenge is to find a reliable audio tagging model which can be trusted as the ground truth.
Yet another direction enabled by this framework is to use these false alarms and misses to automatically correct the caption. If a reliable tagging model can be used as the ground truth, we could develop an audio captioning system which corrects itself while being deployed.


\vfill\pagebreak



\bibliographystyle{IEEEbib}
\bibliography{strings,refs}


\end{document}